\documentclass[reprint,longbibliography,preprintnumbers,nofootinbib,amsmath,amssymb,aps,prd,twocolumns]{revtex4-2}
\usepackage{dblfloatfix}
\usepackage{graphicx}
\usepackage{dcolumn}
\usepackage{bm}
\usepackage{color}
\usepackage{amssymb}
\usepackage{pifont}
\definecolor{green2}{rgb}{0.0, 0.5, 0.0}
\usepackage[dvipsnames]{xcolor}
\usepackage[colorlinks = true,
            linkcolor = blue,
            urlcolor  = blue,
            citecolor = green,
            anchorcolor = blue]{hyperref}
\usepackage{verbatim}
\makeatletter
\usepackage{multirow, graphicx,amssymb,url,mathrsfs,amsmath}
\usepackage{amsxtra,amstext,latexsym,dsfont,amsfonts}
\usepackage{color,eucal}
\usepackage[dvipsnames]{xcolor}
\usepackage{float}
\usepackage{colortbl}
\usepackage{multirow}
\usepackage{tikz}
\usetikzlibrary{tikzmark}
\usetikzlibrary{arrows}

\newcommand{\Rmnum}[1]{\expandafter\@slowromancap\romannumeral #1@}
\usepackage{tabularx, array}
\newcolumntype{L}[1]{>{\raggedright\arraybackslash}p{#1}}
\newcolumntype{C}[1]{>{\centering\arraybackslash}p{#1}}
\newcolumntype{R}[1]{>{\raggedleft\arraybackslash}p{#1}}
\makeatother

\makeatletter
\newcommand{\myBig}{\bBigg@{1.75}}
\makeatother

\begin{document}
\title{Holographic s+p superconductors with axion induced translation symmetry breaking}
\author{Ru-Qing Chen}
\affiliation{Kunming University of Science and Technology, Kunming 650500, China}
\author{Xin Zhao}
\affiliation{Center for gravitation and astrophysics, Kunming University of Science and Technology, Kunming 650500, China and
 College of Physics, Nanjing University of Aeronautics and Astronautics, Nanjing 211106, China}
\author{Hui Zeng}
\email{Corresponding author.}
\affiliation{Kunming University of Science and Technology, Kunming 650500, China}
\author{Zhang-Yu Nie}
\email{Corresponding author.}
\affiliation{Kunming University of Science and Technology, Kunming 650500, China}
\date{\today}
\begin{abstract}
We construct a holographic model for an s+p superconductor with axion-induced translation symmetry breaking within the framework of gauge/gravity duality, working in the probe limit. The equations of motion are solved numerically to investigate the influence of the parameter $k/T$ on the competition and coexistence between the s-wave and p-wave orders. We find that increasing $k/T$ suppresses the thermodynamic stability of both the single condensate s-wave and p-wave solutions. With the $k-\mu$ phase diagram and the condensate curves, we see that the region dominated by the single condensate p-wave phase gradually decreases with the increasing of $k/T$, finally leaving only the single condensate s-wave phase in the large $k/T$ region, which is explained by the grand potential curves showing a slower decreasing of the thermodynamic stability for the s-wave solution than that for the p-wave solution. Furthermore, a larger minimum ratio of the charges $q_p/q_s$ is required to stabilize the s+p coexistent phase as $k/T$ increases, and we determine the precise dependence of this critical ratio on $k/T$. Finally, our study of the optical conductivity reveals that the gap frequency increases with $k/T$. A characteristic kink, associated with the s+p coexistent phase, is identified in the dependence of gap frequency on $k/T$, which could serve as a potential experimental signature for detecting multi-condensate superconductivity.
\end{abstract}
\maketitle

\section{Introduction}\label{Introduction}
The gauge/gravity duality, also known as the AdS/CFT correspondence \cite{Maldacena:1997re}, has emerged as a powerful tool for studying strongly coupled systems, especially after the holographic modeling of superconductors in the HHH model \cite{Gubser:2008px,Hartnoll:2008vx} proposed by Hartnoll, Herzog, and Horowitz. The HHH model has been widely extended and attracted increasing attention, not only shedding light on researches in strongly coupled superconductors but also providing a new perspective for understanding more general phase transitions. In addition to the s-wave holographic superconductor model, p-wave and d-wave orders are also included in the holographic studies \cite{Gubser:2008wv,Chen:2010mk}, which provide new perspectives for understanding the physical properties of the superconductors with different symmetries. Subsequently, the holographic superconductor models with single condensate have been extended to more general cases with multi-condensate. The competition and co-existence of two order parameters are first studied in the holographic s+s models, \cite{Basu:2010fa,Cai:2013wma,Li:2017wbi}, and later extended to the s+p models \cite{Nie:2013sda,Nie:2014qma,Amado:2013lia,Nie:2015zia,Arias:2016nww,Xia:2021pap,Zhang:2023uuq} as well as the s+d models \cite{Nishida:2014lta,Li:2014wca,Liao:2025knf}. These studies reveal that in addition to the superconductor states with single condensate, co-existence between various orders should be considered in a real system, making the phase transitions more abundant and the phase structure more complex.

In the HHH model, even in the normal state, there is always an infinite direct current conductivity. This phenomenon stems from the preservation of translational symmetry in the system. To construct a more realistic holographic superconductor model, it is crucial to introduce momentum relaxation or break the translational symmetry within the holographic model.
There are many ways to realize translation  symmetry breaking in holography. For example, in the holographic scalar lattice\cite{Horowitz:2012ky}, translation  symmetry breaking is achieved by introducing a periodic lattice generated by a source term of a neutral scalar field. In the holographic ionic lattice model\cite{Flauger:2010tv,Horowitz:2012gs}, the lattice is achieved by introducing a spatially varying chemical potential in the boundary theory. The holographic Q-lattice\cite{Donos:2013eha,Donos:2014uba,Ling:2014laa} is realized by introducing a complex scalar field $\phi$ associated with a global U(1) symmetry, which breaks translational invariance and simulate the lattice structure. Massive gravity models\cite{Vegh:2013sk,Davison:2013jba,Blake:2013bqa} introduce mass terms for some gravitons, resulting in the momentum relaxation in boundary field theory. The axion model\cite{Bardoux:2012aw,Andrade:2013gsa,Baggioli:2016oqk} achieves momentum relaxation with (d-1) massless scalar fields, which breaks the translational symmetry.

Against this backdrop, it is important to construct a holographic s+p superconductor model with momentum relaxation to more realistically simulate the physical properties of superconductors with multi-condensates. In current research, we take the axion model and study the competition and coexistence between the s-wave and p-wave orders in the probe limit at different strength of the translational symmetry breaking, which will provide more rich phase structures than the studies with single s-wave or p-wave condensate\cite{Kim:2015dna,Lu:2021tln,Ling:2016lis,Dong:2024fhg,Dong:2025hre}. Although in the probe limit, where the perturbations of the probe fields do not interact with the metric and axion backgrounds, the DC conductivity remains infinite, we are able to get the nontrivial effects from the translational symmetry breaking strength $k/T$ on the phase transitions as well as the energy gap. These results in the probe limit are expected to be the same as the cases with very small back-reaction strength~\cite{Nie:2015zia}. In order to further study the details of the finite DC conductivity, we plan to consider the full back-reaction on the metric and axion fields in the future, based on the useful experiences from this study.

The plan of the article is as follows. In Section~\ref{sec2}~, we establish a holographic s+p superconductor model with translational symmetry breaking in the probe limit. In Section~\ref{sec3}~, we investigate the impact of the translational symmetry breaking strength on phase transitions, presenting the competition and coexistence between the s-wave and p-wave orders. In Section ~\ref{sec4}~, we study the optical conductivity to show the dependence of the energy gap on the translational symmetry breaking strength within this model. In Section~\ref{sec5}~ provides conclusions and discussions.
\section{The holographic model of s+p superconductor with axion induced translational symmetry breaking}\label{sec2}
We consider the following full action of the gravity and matter fields as \cite{Nie:2013sda,Andrade:2013gsa},
\begin{align}
	S=&\,S_{M}+S_{G}~,\label{Lagall}\\
	S_G=&\int d^{4}x\sqrt{-g}(\frac{R}{2}-\Lambda-\frac{X}{2})~,\label{Lagg}\\
       S_M=&\int d^{4}x\sqrt{-g}\Big(-\frac{1}{4}F_{\mu\nu}F^{\mu\nu}
	-D_{\mu}\Psi^{\ast}D^{\mu}\Psi\\
	&\quad\quad  -m_s^{2}\Psi^{\ast}\Psi - \frac{1}{2} \rho_{\mu\nu}^{\dagger} \rho^{\mu\nu}-m_p^2 \rho^\dagger_\mu \rho^\mu\Big)~.\label{Lagm}
    \end{align}

In the gravity part $S_G $,  $\Lambda=-3/L^2$ is the negative cosmological constant where $L$ is the AdS radius, and $X=\frac{1}{2}g^{\mu\nu}\partial_\mu\phi^I\partial_\nu\phi^I$ characterizes the translational breaking sector with the axion fields $\phi^I$. In the matter part $S_M$, $F_{\mu\nu}=\nabla_{\mu}A_{\nu}-\nabla_{\nu}A_{\mu}$ is the Maxwell field strength. $\Psi$ is a complex scalar field, and $\rho_\mu$ is a complex vector field. The covariant derivatives are given by $D_{\mu}\Psi=\nabla_{\mu}\Psi-i q _sA_\mu\Psi$ and $\rho_{\mu\nu}=\bar{D}_\mu \rho_\nu - \bar{D}_\nu \rho_\mu$, where $\bar{D}_\mu \rho_{\nu}=\partial_\mu\rho_{\nu}-i q_p A_\mu\rho_{\nu}$. Here, $q_s$, $q_p$ and $m_s$, $m_p$ are the charges and masses of $\Psi$ and $\rho_\mu$, respectively.

The gravity part $S_g$ admits an asymptotically AdS black hole with a planar horizon topology, as described in the following line element
\begin{align}
	&\  ds^2 = -f(r)dt^2+\frac{dr^2}{f(r)}+r^2(dx^2+dy^2)~,
\end{align}
    where
\begin{align}
	&\ f(r)= \frac{r^2}{L^2} \left(1-\frac{r_h^3}{r^3}\right)- \frac{k^2}{2L^2}\left(1-\frac{r_h}{r}\right)~.
\end{align}

In this metric, $r_h$ denotes the location of the event horizon, where $f(r_h)=0$. The axion fields are given by $\phi^I=kx^I$,  where $x^1=x$ and $x^2 =y$. Here, $k$ can be interpreted as an axion parameter. When $k \to 0$, the system returns to the Schwarzschild-AdS scenario. Meanwhile, the Hawking temperature is given by
\begin{align}
    T= \frac{1}{4\pi L^2}(3r_h-\frac{k^2}{2r_h})~.
\end{align}

We set the following ansatz for the matter fields
\begin{align}\label{ansatz}
\quad A_t=\Phi(r), \quad\Psi=\Psi_s(r), \quad\rho_x=\Psi_p(r) ~,
\end{align}
and all other field components are set to zero. Consequently, the equations of motion for the matter fields are as follows
\begin{align}
    \frac{2 q_p^2 \Phi \Psi_p^2}{r^2 f} + \frac{2 q_s^2 \Phi \Psi_s^2}{f} - \frac{2 \Phi'}{r} -\Phi'' = 0~\label{equation9},\\
    \frac{m_s^2 \Psi_s}{f} - \frac{q_s^2 \Phi^2 \Psi_s}{f^2} - \frac{2 \Psi_s'}{r} - \frac{f' \Psi_s'}{f} - \Psi_s'' = 0~\label{equation10},\\
    \frac{m_p^2 \Psi_p}{f} - \frac{q_p^2 \Phi^2 \Psi_p}{f^2} - \frac{f' \Psi_p'}{f} - \Psi_p'' = 0~\label{equation11}.
\end{align}

To solve these coupled equations, we need to specify the boundary conditions. The expansions near the horizon $r=r_{h}$ are
\begin{align}
\Phi(r) &= \Phi_1(r - r_h) + \mathcal{O}(r - r_h)^2~,  \\
\Psi_s(r) &= \Psi_{s0} + \Psi_{s1}(r - r_h) + \mathcal{O}(r - r_h)^2~, \\
\Psi_p(r) &= \Psi_{p0} + \Psi_{p1}(r - r_h) +\mathcal{O}(r - r_h)^2~.
\end{align}

The value of $\Phi(r=r_h)$ is set to zero to ensure the finiteness of $g^{\mu\nu}A_{\mu}A_{\nu}$, and the asymptotic expansions near the boundary $r\to\infty$ are given by
\begin{align}
\Phi(r) &= \mu - \frac{\rho}{r} + \ldots~,  \\
\Psi_s(r) &= \frac{\Psi_{s-}}{r^{\Delta_{s^-}}} + \frac{\Psi_{s+}}{r^{\Delta_{s^+}}} + \ldots~,\\
\Psi_p (r)&= \frac{\Psi_{p-}}{r^{\Delta_{p^-}}} + \frac{\Psi_{p+}}{r^{\Delta_{p^+}}} + \ldots~,
\end{align}
where
\begin{align}
\Delta_{s\pm} &= \frac{3\pm \sqrt{9 + 4m_s^2}}{2}~,  \\
\Delta_{p\pm} &= \frac{1 \pm \sqrt{1 + 4m_p^2}}{2}~.
\end{align}

Due to the AdS/CFT dictionary, $\mu$ and $\rho$ are the chemical potential and charge density of the boundary system, respectively. We adopt the standard quantization, in which $\Psi_{s-}$ and $\Psi_{p-}$ are considered as the source terms of the boundary operators, while $\Psi_{s+}$ and $\Psi_{p+}$ are regarded as the vacuum expectation values. We set the conditions $\Psi_{s-} = \Psi_{p-} = 0$ to obtain the solutions for spontaneous U(1) symmetry breaking.

In this paper, we conduct our research within the grand canonical ensemble. To compare the stability of different solutions, we calculate the grand potential of the various solutions. In the probe limit, the contribution from the gravity part are the same for different solutions. Therefore the difference in the grand potential originates solely from the matter part of the action, although it is infinitesimal compared with the gravity part. After substituting the equations of motion into the matter part of the Euclidean action, the contribution to the grand potential from the matter part is 
\begin{align}
	\Omega_m=\frac{V_2 }{T}(-\frac{\mu\rho }{2}-\int_{r_h}^{\infty}(\frac{q_p^2 \Phi^2 \Psi_p^2}{f}+\frac{q_s^2 r^2\Phi^2 \Psi_s^2}{f})dr)~.
\end{align}
The equations of motion (\ref{equation9},\ref{equation10},\ref{equation11}) exhibit the following three sets of scaling symmetries, which are useful to simplify the numerical procedures.
\begin{align}
(\text{I})~ &\Phi \rightarrow \lambda^{-1} \phi, \Psi_p \rightarrow \lambda^{-1}\Psi_p, f \rightarrow \lambda^{-2}f, r \rightarrow \lambda^{-1}r, \nonumber\\
&k \rightarrow \lambda^{-1}k, r_h \rightarrow \lambda^{-1}r_h~; 
\label{scaling1}\\
(\text{II})~ &\Phi \rightarrow \lambda^{2} \phi, \Psi_p \rightarrow \lambda\Psi_p,  \Psi_s \rightarrow \lambda\Psi_s,f \rightarrow \lambda^{2}f,  \nonumber\\
&m_s^2\rightarrow \lambda^{-2}m_s^2,m_p^2\rightarrow \lambda^{-2}m_p^2,  L \rightarrow \lambda^{-1}L~; 
\label{scaling2}\\
(\text{III})~ &\Phi \rightarrow \lambda \phi, \Psi_p \rightarrow \lambda\Psi_p,\Psi_s \rightarrow \lambda\Psi_s,  \nonumber\\
&\ q_s \rightarrow \lambda^{-1} q_s, \ q_p \rightarrow \lambda^{-1} q_p~.
\label{scaling3}
\end{align}

Regarding the aforementioned scaling symmetries, we set $r_h=L=1$ for numerical calculations. After obtaining the numerical solutions, we can use the two sets of scaling symmetries (\ref{scaling1},\ref{scaling2}) to revert $r_h$ and $L$ to any value. Due to the scaling symmetry (\ref{scaling3}), only the ratio $q_p/q_s$ is important in tuning the phase transitions, therefore $q_s$ is set to 1 without lose of generality.
\section{The phase transitions involving the s-wave and p-wave orders}\label{sec3}
\subsection{The single condensate s-wave and p-wave solutions}\label{sec3.1}
In this system, there are three distinct solutions with  non-zero condensates: the s-wave solution, the p-wave solution, and the s+p coexistent solution. By turning off either the p-wave or s-wave order, we can readily obtain the s-wave solution or the p-wave solution. To focus on the effects of the translational symmetry breaking strength $k/T$, we fix the mass parameters as $m_s^2=0$ and $m_p^2 = 3/4$ throughout the remainder of this paper.
\begin{figure}[h]
    \includegraphics[width=0.49\columnwidth]{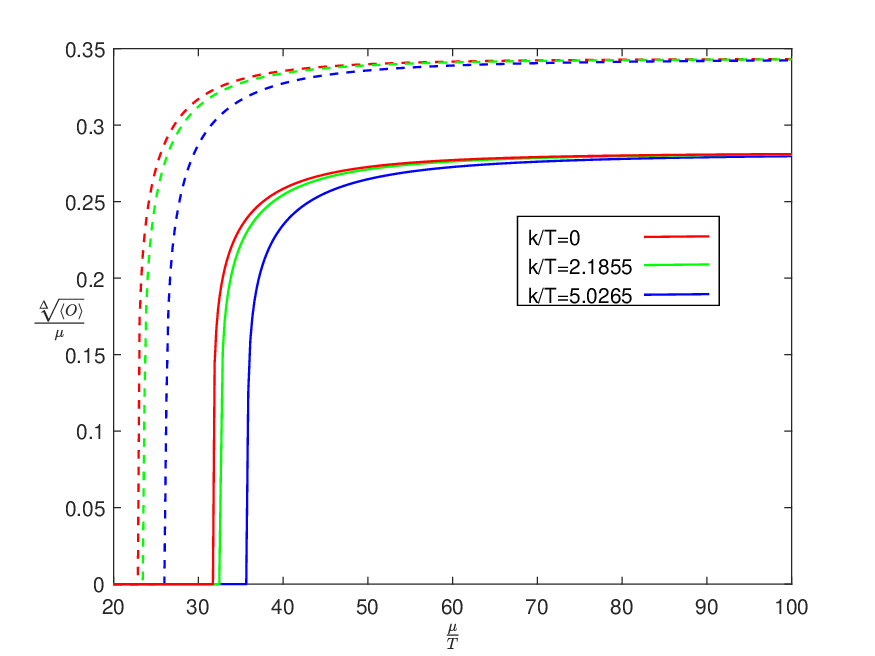}
    \includegraphics[width=0.49\columnwidth]{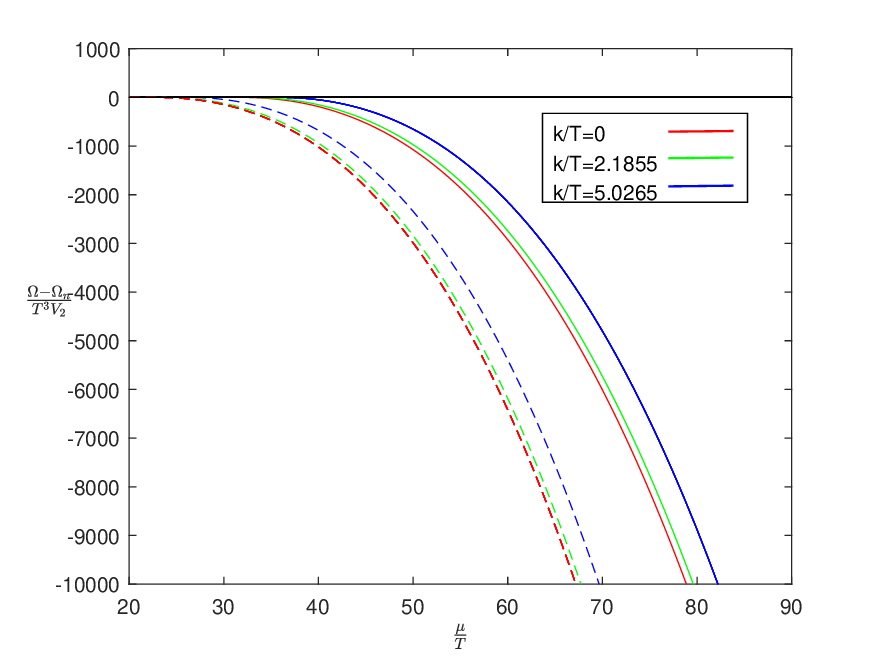}
	\caption{The condensate (Left) and grand potential (Right) curves of the s-wave (Solid Curves)  and p-wave solutions (Dashed Curves) with $q_s=q_p=1$ and three different values of the translational symmetry breaking strength ($k/T$=0 (red), 2.1855 (green), and 5.0265 (blue)).}\label{1}
\end{figure}

In Figure~\ref{1}, we plot the condensate and grand potential curves of the single condensate s-wave and p-wave solutions with $q_s=q_p=1$ and three different values of the translational symmetry breaking strength $k/T$. We can see that as $k/T$ increases, the single condensate s-wave and p-wave solutions exhibit the same qualitative law: their critical values of chemical potential $\mu_c/T$ both increase, and their grand potential curves both rise up, indicating a reduction in the overall stability of the single condensate solutions. While at the same time, the condensate value of both the s-wave and p-wave solutions tend to the respective constant values at very low temperatures, which is insensitive to the change of $k/T$.
\subsection{The multi-condensate s+p solutions}\label{sec3.2}
\begin{figure}
	\includegraphics[width=0.89\columnwidth]{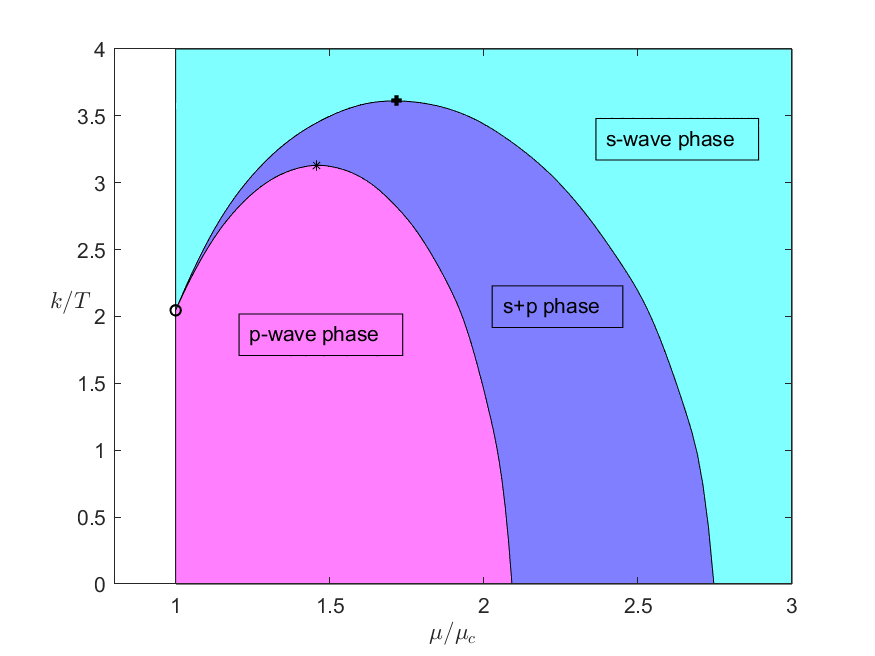}
	\caption{The $k-\mu$ phase diagram with $q_p=0.7225$ and $q_s=1$. The white region is dominated by the normal phase. The cyan region is dominated by the single condensate s-wave phase and the magenta region is dominated by the single condensate p-wave phase. The blue region is dominated by the multi condensate s+p phase. The symbols \textit{o}, \textit{*}, \textit{+} mark the three special points and their values of $k/T$ are given by $(k/T)^o=2.0461$, $(k/T)^*=3.1311$ and $(k/T )^+=3.6118$, respectively.}\label{2}
\end{figure}

In the following analysis of the s+p coexistent phase, we fix $q_s=1$ without lose of generality. We first choose an appropriate value for $q_p/q_s=q_p$ in order to present the s+p coexistent phase at $k/T=0$. Due to previous studies, the coexistent phase usually show up near the intersection point of the grand potential curves of the s-wave and p-wave solutions. Thus we tune the value of $q_p$ to shift the grand potential curve of the p-wave solution "parallel"\cite{Li:2017wbi} and reach a good choice $q_p=0.7225$, where the grand potential curve of the p-wave solution intersect with the grand potential curve of the s-wave solution. Next we fix this value of $q_p$ and study the effect of the translational symmetry breaking strength $k/T$ on the competition and coexistence between the s-wave and p-wave orders. 
In Figure~\ref{2}, we plot the phase diagram of $k/T$ versus $\mu/\mu_c$ to present the results, where four distinct regions are observed: the normal phase (white), the s-wave phase (cyan), the p-wave phase (magenta) and the s+p coexistent phase (blue). The boundary lines between these regions represent the critical points of second-order phase transitions. As $k/T$ increases, the interval dominated by the single condensate p-wave phase gradually decreases until $(k/T)^*$. Above $(k/T)^*$ the p-wave order only exists in the s+p coexisting phase, and finally disappears in the region $k/T>(k/T)^+$. We can also see a quadruple point at $k/T=(k/T)^o$. Since the results of single condensate solutions show that the stability of both the s-wave and p-wave solutions are reduced along with the increasing of $k/T$, this phase diagram further indicate that the reduction of the stability for the single condensate s-wave solution is slower than that for the single condensate p-wave solution, which need to be confirmed by the results from the grand potential curves.

\begin{figure}
	\includegraphics[width=0.49\columnwidth]{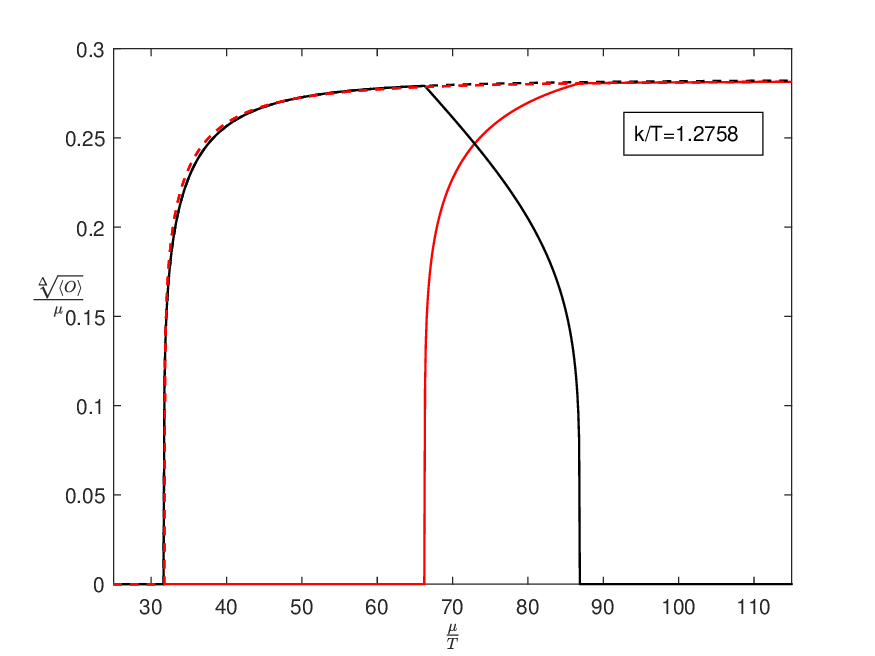}
    \includegraphics[width=0.49\columnwidth]{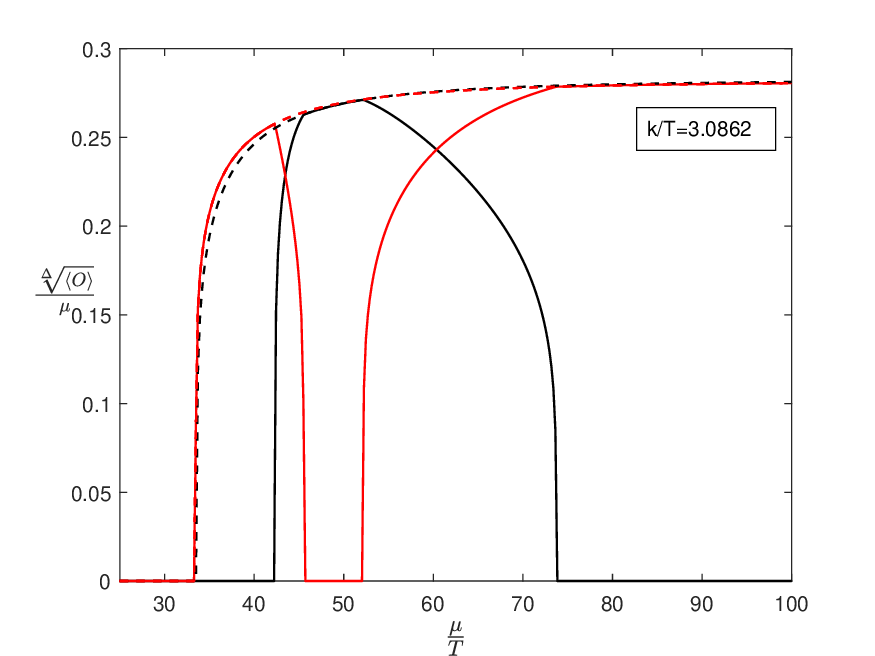}
	\includegraphics[width=0.49\columnwidth]{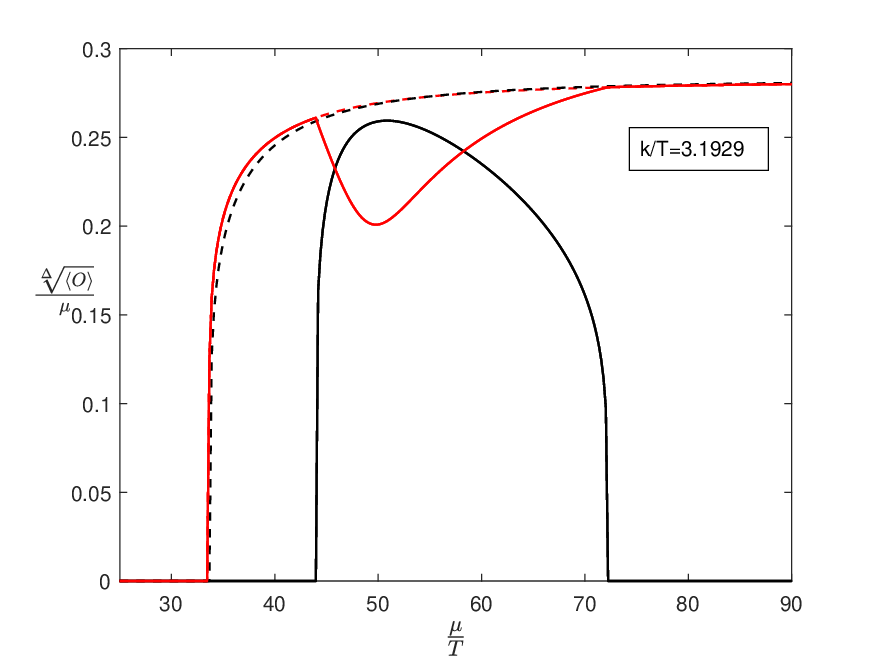}
    \includegraphics[width=0.49\columnwidth]{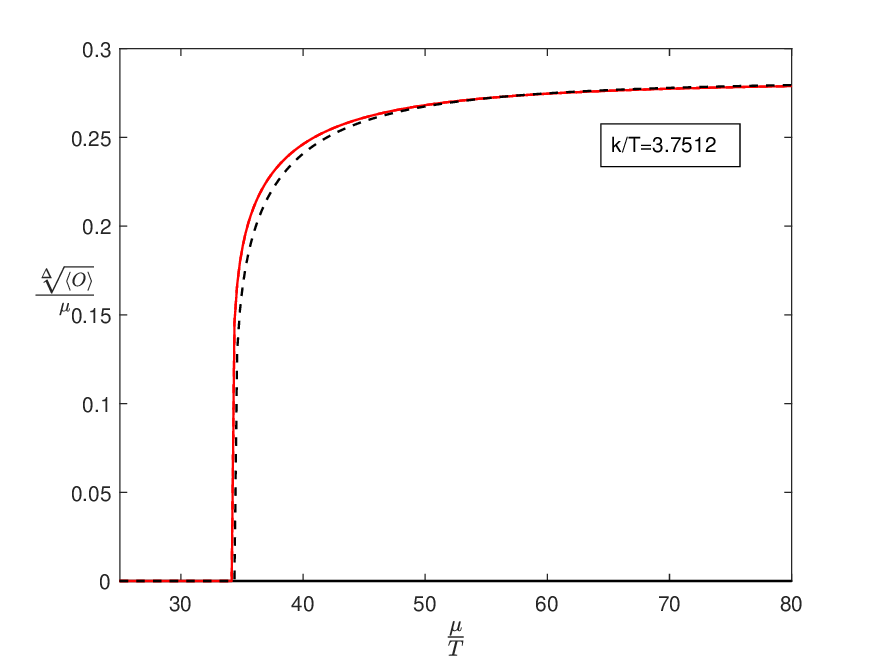}
	\caption{The condensate curves for various fixed values of the translational symmetry breaking strength $k/T =1.2758, 3.0862, 3.1929$, and $3.7512$. The red and black lines indicate the condensate values of the s-wave and p-wave orders, respectively. Solid lines denote the condensate values of the most stable solutions, while dashed lines denote the condensate values in the unstable sections of the single condensate solutions.
}\label{3}
\end{figure}

In Figure~\ref{3} , we present the detailed condensation curves at four values of $k/T$ to present the typical cases. We can see that with $k/T=1.2758$, as the chemical potential $\mu/T$ gradually increases from the position dominated by the normal phase, the p-wave order condensates before the s-wave order. Later the condensate of the s-wave order grows up in the multi-condensate s+p phase along with the decreasing of the p-wave condensate, presenting a typical X-type phase transition. Finally the p-wave condensate vanishes at the last critical point and the system is dominated by the single condensate s-wave phase in the region of large chemical potential $\mu/T$. When $k/T$ is larger than $(k/T)^o$, the s-wave order condensates before the p-wave order along the increasing of $\mu/T$, as shown in the last three plots in Figure~\ref{3}. The second plot in Figure~\ref{3} show the condensate curves with $k/T=3.0862$, where we see that with the increasing of the chemical potential $\mu/T$, the s-wave order condensates first. Then the p-wave order grows up in the multi-condensate s+p phase along with the decreasing of the s-wave condensate and goes into the single condensate p-wave phase. However, later the condensate of the s-wave order grows up again in the second section of the multi-condensate s+p phase along with the decreasing of the p-wave condensate. Finally the p-wave condensate vanishes and the single condensate s-wave phase dominate in the region of large chemical potential $\mu/T$. The third plot in Figure~{3} present the condensate curves with $k/T=3.1929$, where the feature is that the condensate of the p-wave order grows up first and later decreases in the single section of the multi-condensate s+p phase, presenting a typical n-type condensate curve and a reentrance to the single condensate s-wave phase. In the last panel of Figure~\ref{3}, we show the condensate curves with $k/T=3.7512$, where the system is always dominated by the single condensate s-wave phase.

\begin{figure}
	\includegraphics[width=0.89\columnwidth]{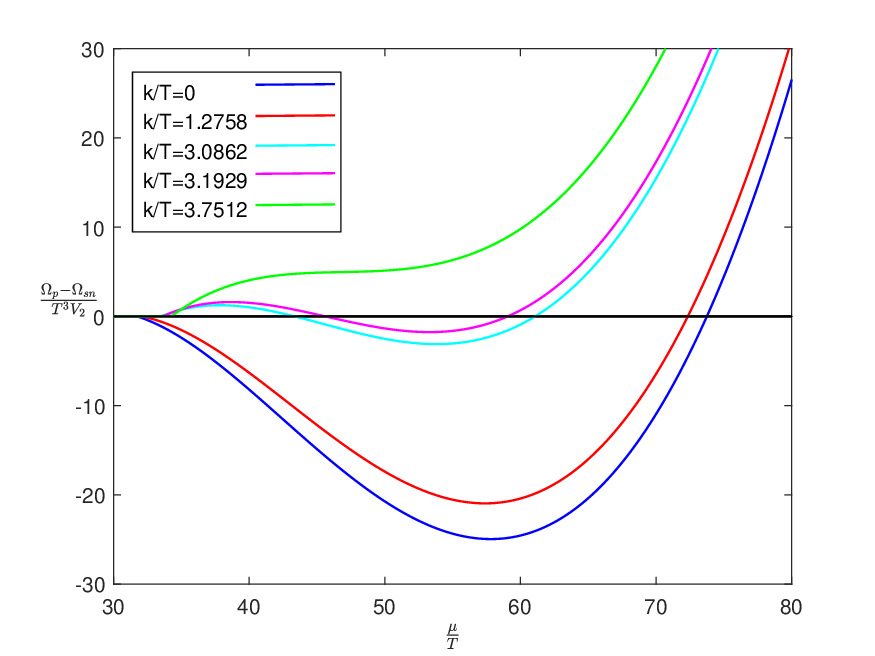}
	\caption{The grand potential curves for $k/T=0$ (blue), $1.2758$ (red), $3.0862$ (cyan), $3.1929$ (magenta), and $3.7512$ (green), respectively. The relative value of the grand potential for the single condensate p-wave solution with respective to the single condensate s-wave solution (or normal solution below the critical point of the s-wave solution) is presented.
    }\label{4}
\end{figure}

In order to better understand the above phase transitions, we plot the relative value of the grand potential of the single condensate p-wave solutions with respect to the single condensate s-wave solution $\Omega_p-\Omega_{s}$ with five different values of $k/T$ in Figure~\ref{4}. In the cases that the p-wave order condenses before the s-wave order, we plot the relative value with respect to the normal phase instead in the region below the critical chemical potential of the single condensate s-wave order. Therefore we use a combined notation $\Omega_p-\Omega_{sn}$ to denote this relative value of grand potential. The curves in  Figure~\ref{4} shows that as $k/T$ increases, the grand potential curve for the relative value of the p-wave solution is shifted upward, and the region where the p-wave solution get the lowest grand potential density shrinks accordingly. These results show that with the increasing of $k/T$, the thermodynamic stability of the single condensate s-wave solution decreases slower than the p-wave solution, therefore the p-wave order gradually loses the competition against the s-wave order.

\begin{figure}[h]
	\includegraphics[width=0.33\columnwidth]{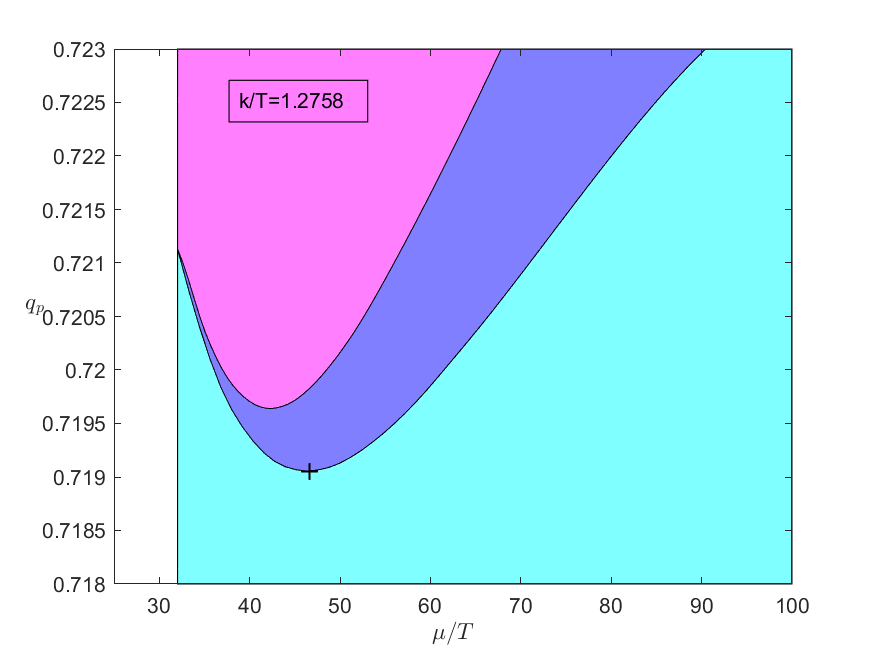}
    \includegraphics[width=0.33\columnwidth]{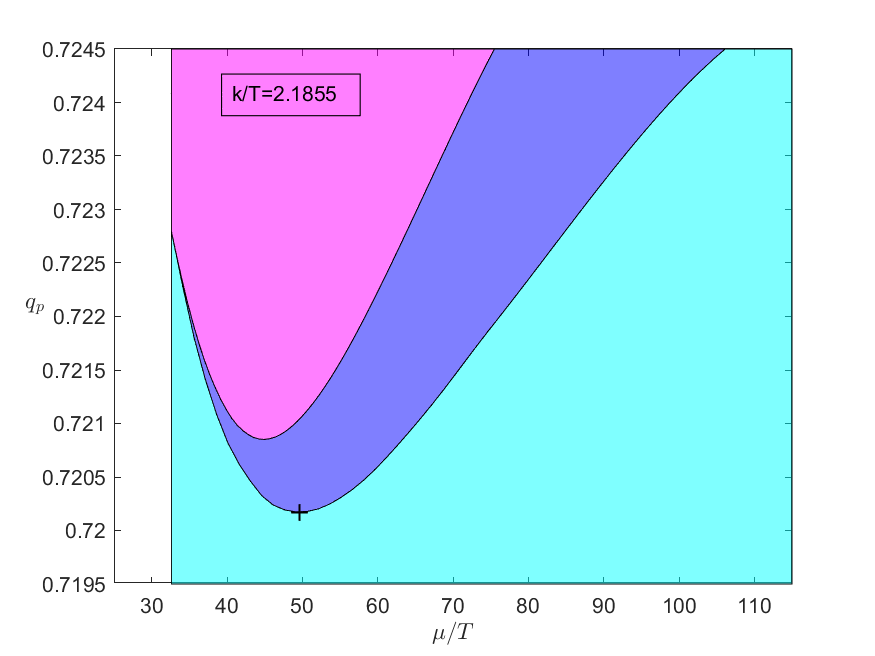}
      \includegraphics[width=0.31\columnwidth]{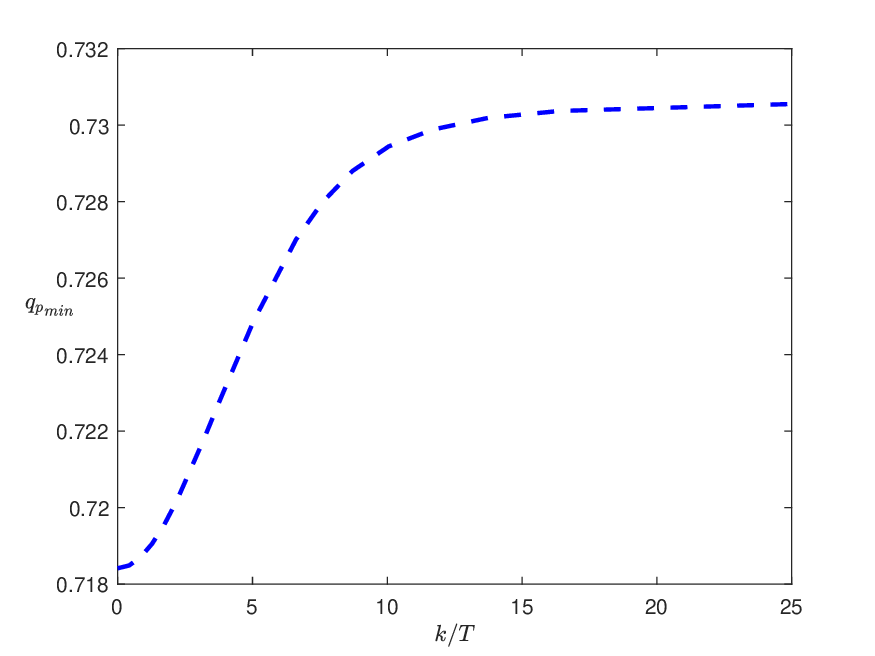}
	\caption{The $q_p-\mu$ phase diagrams with $k/T=1.2758$ (Left) and $k/T = 2.1855$ (Middle), and the dependence of $q_{min}$ on $k/T$ (Right). In the left and middle panels, the white region represents the normal phase, the cyan region represents the s-wave phase, the magenta region represents the p-wave phase and the blue region represents the s+p phase. The symbol ``+'' signifies the minimum charge of the p-wave order $q_p=q_p/q_s=q_{Pmin}$ required for the existence of the multi condensate s+p solutions, the dependence of which on $k/T$ is presented in the right panel.}
    \label{5}    
\end{figure}

It is also necessary to consider general values of the ratio $q_p/q_s=q_p$. We set two values of $k/T$ and plot the $q_p-\mu$ phase diagrams in the left and middle plots of Figure~\ref{5}. We can see that the phase diagrams are also divided into four region dominated by the normal phase, the s-wave phase, the p-wave phase and the s+p phase, respectively, and the topologies of the phase diagrams are the same. From these phase diagrams, we see that even at a fixed value of $k/T$, it is always possible to get the various phase transitions as shown in Figure~\ref{3} by tuning the value of $q_p$. At a fixed value of $k/T$, there is a minimal value of $q_p$, below which the p-wave condensate is always zero and the s+p coexistent solution do not show up. This minimal value ${q_p}_{min}$ depends on $k/T$, and we show the dependence in the right plot of Figure~\ref{5}, where we can see that ${q_p}_{min}$ increases with the increasing of $k/T$, and approaches a constant value at large $k/T$. This result suggests that with a larger value of the translational symmetry breaking strength $k/T$, a larger $q_p$ value is necessary for the condensate of the p-wave order as well as the emergence of the multi-condensate s+p solution.
\section{The optical conductivity and the energy gap}  \label{sec4}
In this section, we investigate the optical conductivity of the holographic s+p superconductor with the impact of the translational symmetry breaking strength. We consider the linear perturbations of the gauge field component as $\delta A_y=A_y(r)e^{-i\omega t}$ with the perturbations for other components set to zero. In principle, we should also consider the perturbations of the $A_x$ component, especially in the anisotropic p-wave and s+p phases. However, the perturbations of the $A_x$ component will couple to the background field $\rho_x$ and involve in other constrain equations. In this work, we focus on the effects of the translational symmetry breaking strength, and therefore only consider the simplest case of conductivity $\sigma_{yy}$ by investigating the perturbation of the $A_y$ component. The decoupled equation for the perturbation $A_y$ is
\begin{align}
(\frac{\omega^2}{f^2} - \frac{2 q_p^2 \Psi_p^2}{r^2 f} - \frac{2 q_s^2  \Psi_s^2}{f} )A_y+ \frac{ f'}{f}A_y' + A_y''=0~.
\end{align}
We apply the ingoing boundary condition at the event horizon
\begin{align}
 A_y(r) \rightarrow (r - r_h)^{-i\omega/(3r_h - \frac{k^2}{2r_h})}(1+ \cdots)~,
 \end{align}
while the asymptotic behavior of \(A_y\) at large radius is
\begin{align}
	A_y(r) \rightarrow A^{(0)} + \frac{A^{(1)}}{r} + \cdots~.
\end{align}
The AdS/CFT correspondence tells us that the dual source and the current expectation value are represented by \(A^{(0)}\) and \(A^{(1)}\), respectively. Therefore, according to Ohm's law, we obtain the conductivity as
\begin{align}
\sigma_{yy}(\omega) = - \frac{iA^{(1)}}{\omega A^{(0)}}~.
\end{align}

\begin{figure}
\includegraphics[width=0.49\columnwidth]{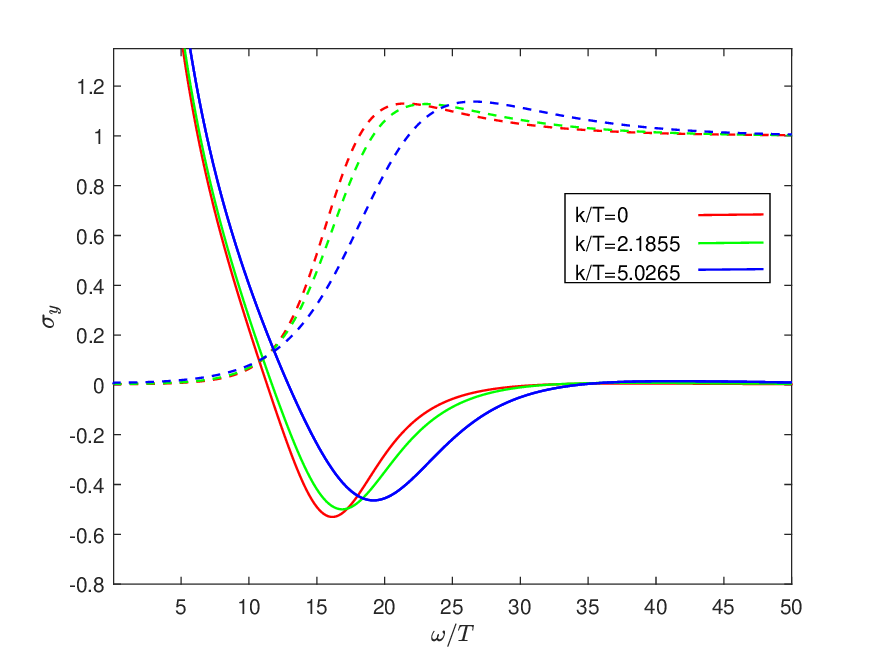}
\includegraphics[width=0.49\columnwidth]{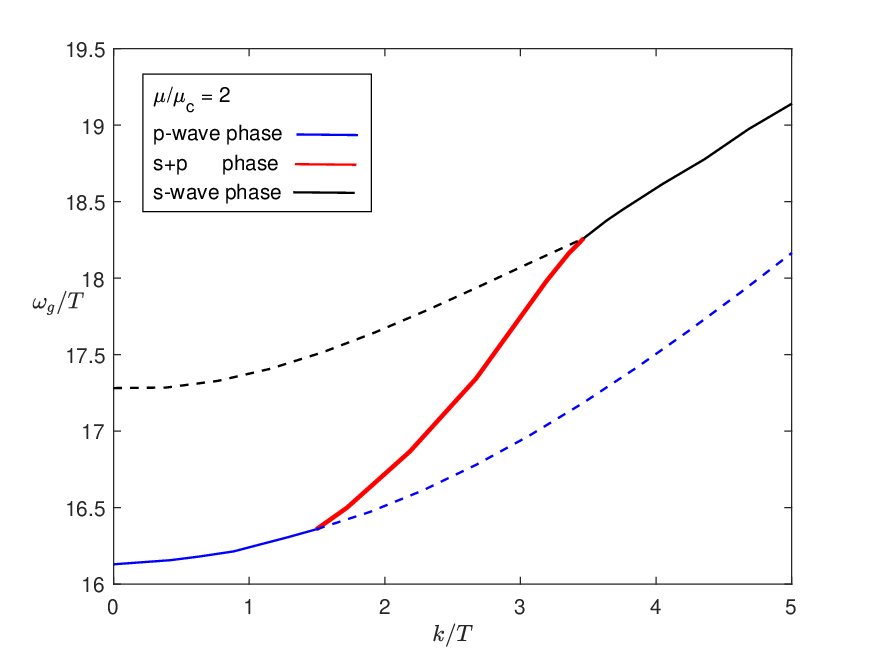}
	\caption{\textbf{Left:} The real part (dashed lines) and imaginary part (solid lines) of the optical conductivity for various values of $k/T$ with $\mu/\mu_c=2$ and $q_p = 0.7225$. \textbf{Right:} The dependence of the energy gap $\omega_g/T$ on to $k/T$ with $\mu/\mu_c=2$ and $q_p = 0.7225$. The black, blue and red lines mark the results for the s-wave, p-wave and s+p solutions, respectively. The solid lines indicate stable sections for the single condensate solutions while the dashed lines indicate unstable ones.
}\label{6}
\end{figure}

With the above schedule, we obtain the optical conductivity on the s-wave, p-wave and s+p superconductor phases. We still take $q_p=0.7225$ and fix $\mu/\mu_c=2$ to focus on the influence of the translational symmetry breaking strength $k/T$. In the left panel of Figure~\ref{6}, we choose three different values of the translational symmetry breaking strength $k/T$,  which in the phase diagram corresponds to the s-wave phase, the p-wave phase and the s+p phase, respectively, and plot the real and imaginary parts of the optical conductivity. In this plot, we can see the typical frequency gap $\omega_g$ in the three different superconductor phases. The accurate position of the frequency gap is able to be located by minimum of the imaginary part $Im(\sigma_y)$. In addition, the frequency gap seems to increase along with the increasing of $k/T$. Therefore, to better present the dependence of the frequency gap $\omega_g$ on the translational symmetry breaking strength $k/T$, we plot this curve in the right panel of Figure~\ref{6}. It is clear that as $k/T$ increases, the frequency gap $\omega_g/T$ increases monotonically, which is consistent with the effect of translational symmetry breaking on the single condensate p-wave holographic superconductors reported in reference\cite{Lu:2021tln,Dong:2024fhg}. Furthermore, the curve exhibit a kink due to the coexistent s+p phase bridging the single condensate s-wave and p-wave phases. This feature may serve as a useful signal for the experimental detection of such multi-condensate phases.
We should also notice from the left panel of Figure~\ref{6} that the imaginary part of the conductivity $\text{Im}[\sigma]$ exhibits a pole near the zero-frequency. According to the Kramers-Kronig relations, this pole implies a Dirac delta function in the real part $\text{Re}[\sigma]$ at zero frequency, indicating an infinite direct current (DC) conductivity, which is a characteristic signal of superconductors. However, this infinite DC conductivity also exist in the normal phase due to the translational symmetry, which is expected to be deformed to a finite DC conductivity with considering the translational symmetry breaking with a nonzero $k/T$. However, in this study, we take the probe limit, which cut off the connection between the perturbations of metric and matter fields. As a result, the normal phase still get an infinite DC conductivity.

\section{Conclusions and discussions}\label{sec5}
This article explores a holographic s+p superconductor model with translational symmetry breaking using the gauge/gravity duality. The study focuses on the effects of the translational symmetry breaking strength $k/T$ on the competition and coexistence between the s-wave and p-wave orders, as well as the optical conductivity.

For the single condensate s-wave and p-wave solutions, the critical value of $\mu_c/T$ increases along with the increasing of the translational symmetry breaking strength $k/T$. The grand potential curves of both the single condensate s-wave and p-wave solutions rise up as the translational symmetry breaking strength $k/T$ increases. These results indicate that a larger $k/T$ reduces the stability of the single condensate s-wave and p-wave solutions.

We plot the $k-\mu$ phase diagram including the multi condensate s+p phase with $q_p=0.7225$ to show the phase structure, and present the detailed condensate curves in four typical cases. We can see that along with the increasing of chemical potential $\mu/T$, at $k/T=0$, the p-wave order condensate first and the system exhibits a typical X-type phase transition. As $k/T$ increases to be larger than the special value $(k/T ) ^o=2.0461$, the s-wave order condensate first instead, and two sections of the s+p phase with the X-type condensate curves are observed and the p-wave order only exist in the middle region. With an larger value of $k/T$, the two sections of the s+p phase merge into one section where the p-wave condensate increase at first and then decrease, showing the reentrance back to the s-wave phase with the n-type condensate curve for the p-wave order. With further increasing of $k/T$, The region of the s+p phase also diminishes gradually, and finally only the single condensate s-wave phase is left in the phase diagram with a large value of $k/T$. The phase structure and condensate curves are well explained by the grand potential curves, which show that with the increasing of $k/T$, the thermodynamic stability of the single condensate s-wave solution decreases slower than the p-wave solution, and the p-wave order gradually loses the competition against the s-wave order.

The $q_p-\mu$ phase diagram also include the four different phases, and indicate a minimum value $q_{min}$ for the p-wave order as well as the s+p coexistent phase. With a greater translational symmetry breaking strength, the value of $q_{min}$ becomes larger, indicating that a larger value of $q_p$ is required for the p-wave condensate as well as the coexistence between the s-wave and p-wave orders.

The results of optical conductivity show typical frequency gap and infinite DC conductivity for the s-wave, p-wave and s+p superconductor phases. The frequency gap increase monotonically along with the increasing of $k/T$, and show a kink due to the coexistent s+p phase, which may serve as a useful signal for the experimental detection of such multi-condensate phases.

In this study, we only consider the probe limit to focus on the influence of $k/T$ on the phase transitions and energy gap. This preserves the infinite DC conductivity in the normal phase. It is important to include the back-reaction on the metric to analyze the behavior of DC conductivity. It is also interesting to include the perturbations of $A_x$ component to study the anisotropic behavior of the conductivity in the p-wave and s+p phases. We expect these problems to be solved in future studies.
\section*{Acknowledgments}
This work was supported by the National Natural Science Foundation of China (Grant Nos. 11965013 and 12575054). ZYN is partially supported by Yunnan High-level Talent Training Support Plan Young \& Elite Talents Project (Grant No. YNWR-QNBJ-2018-181).
\bibliographystyle{apsrev4-1}
\bibliography{reference}
\end{document}